\documentclass[apj]{emulateapj}

\shorttitle{Eigen-reconstruction of Tensor Perturbations}
\shortauthors{M.~ Farhang, A.~ Vafaei Sadr}
\usepackage{color}
\usepackage[colorlinks=true,
            linkcolor=black,
            urlcolor=black,
            citecolor=blue, breaklinks=true]{hyperref}
\usepackage{url}
\usepackage{amsmath}
\usepackage{amssymb}
\usepackage{gensymb}

\usepackage{graphics}
\usepackage{graphicx}
\usepackage{natbib}
\usepackage{psfig}
\input{epsf}
\usepackage{ae,aecompl}
\usepackage{booktabs}

\newcommand{\be}{\begin{equation}}
\newcommand{\ee}{\end{equation}}

\begin{document}

\title{Eigen-reconstruction of Perturbations to the Primordial Tensor Power Spectrum}

\author{M. Farhang\altaffilmark{1}, A. Vafaei Sadr\altaffilmark{1,2}}
\affil{\scriptsize
$^{1}${Department of Physics, Shahid Beheshti University, Velenjak, Tehran 19839, Iran}\\
$^{2}$ School of Physics, Institute for Research in Fundamental Sciences (IPM), P. O. Box 19395-5531, Tehran, Iran}
\email{Corresponding author email: m\_farhang@sbu.ac.ir}

\begin{abstract}

We explore the potential of the  B-mode anisotropies of the Cosmic Microwave Background radiation (CMB) to constrain the shape of the primordial tensor power spectrum in a model-independent way.
We expand  possible perturbations to the  power-law primordial tensor  spectrum (predicted by the simplest single-field slow-roll inflationary models) using various sets of localized and nonlocalized basis functions and construct the Fisher matrix for their amplitudes. 
The eigen-analysis of the Fisher matrix would  then yield a hierarchy of uncorrelated perturbation patterns  (called tensor eigenmodes or TeMs) which are rank-ordered according to their measurability by data. 
We find that the first three TeMs are expected to be constrainable within a few percent by the next generation of B-mode experiments. 
We discuss how the method can be iteratively used to reconstruct the observable part of any general deviation from the fiducial power spectrum. 

 \end{abstract}


\section{Introduction}
The anisotropies of the CMB B-mode polarization at large angular scales, sourced by primordial tensor fluctuations (aka gravitational waves), is a unique window to the yet unknown physics of the very early universe.
Many experiments (e.g., see \citep{mat14,del18,kog16}) are being designed or are under construction to observe this imprint of gravitational waves. 

 The primordial power spectrum of tensor fluctuations, $P_{\rm t}(k)$,  is often approximated by a power-law, as predicted by the simplest single-field slow-roll inflationary models,
%---------------------------------------------------------------------------------------
\begin{equation}\label{fiducial}
P_{\rm t}(k) \approx A_{\rm t}(k_{\rm t})(k/k_{\rm t})^{n_{\rm t} (k_{\rm t})}.
\end{equation}
%---------------------------------------------------------------------------------------
This power spectrum is also equivalently parametrized by $r$ (and $n_{\rm t}$), where $r \equiv P_{\rm t}(k_{\rm t})/P_{\rm s}(k_{\rm t})$ is the tensor-to-scalar ratio, $P_{\rm s}$ is the primordial scalar power spectrum and $k_{\rm t}$ is the tensor pivot scale.
Near-future B-mode experiments aim to measure $r$, and potentially $n_{\rm t}$ by cosmic variance limited measurements of B-mode fluctuations. 

More general models of early Universe predict power spectra that could differ from this simplest shape.
In this work, we assess the observability of  possible deviations from the power-law shape  in a semi-blind, or model-independent, way. 
The most observable shapes of these deviations are the patterns best constrained by data, and are called tensor eigenmodes, or TeMs. 
In Section~\ref{sec:TeMs}  we discuss the methodology to generate these TeMs, investigate their various properties such as convergence and robustness against different assumptions used in their construction, and study their sensitivity to observational assumptions, as well as the fiducial model.
We also explore how these modes can be used as a
 practically complete basis for the expansion of  any general form of fluctuations $\delta \ln P(k)$ around the fiducial power spectrum  (Eq.~\ref{fiducial}).
  We conclude by a detailed discussion of how one could iteratively proceed toward an optimal and blind extraction of the available information about the physics of early Universe from the B-mode spectrum (Section~\ref{sec:disc}).

%--------------------------------------------------------------------------------------------------------------------
\section{Tensor Eignemodes}\label{sec:TeMs}
%--------------------------------------------------------------------------------------------------------------------
In this section we aim to go beyond
the standard practice in CMB B-mode analysis and search for potential deviations from the power-law 
approximation of the primordial tensor power spectrum  (Eq.~\ref{fiducial}).
 We choose a semi-blind approach, as explained in detail below, to avoid biases from theoretically-motivated
  scenarios and learn about possible deviations as is preferred by observations.

 The B-mode power spectrum was recently used to investigate the reconstruction of perturbations to the primordial power-law  spectrum by expanding possible deviations in top-hot bases and forecasts were made for the uncertainties in their amplitudes given various observational scenarios \citep{hir18}. 
 Here, however, we focus on the eigen-reconstruction of tensor perturbations. This way, instead of searching for all possible patterns of deviations in the tensor spectrum, our search  will be limited to the patterns most preferred by data and therefor are best constrained. 
 This data-driven mode reconstruction would enormously optimize the searching pipeline for deviations  and the associated parameter estimation procedure, as we will see below. Eigen-analysis (also known as principal component analysis) has been earlier used in various fields of cosmological data analysis (originally used to construct parameter eigenmodes in CMB data analysis in \citep{bond97}, and later in various parameter estimation problems such as studying different reionization scenarios in \citep{hu03} and \citep{mor08}). 

 Here,  we closely follow the notation of \cite{far12} where the parameter eigenmode approach was used to investigate the observability of possible deviations from the standard recombination scenario.

 In the following, we first introduce the methodology to construct the eigenmodes  (Section~\ref{sec:method}),
  then present the eigenmodes constructed with different assumptions (Section~\ref{sec:tem}) and illustrate how the modes can be used as a complete basis to expand fluctuations around the fiducial scenario (Section~\ref{sec:models}).
%-----------------------------------------------------------------------------------------------
\subsection{Methodology}\label{sec:method}
Our method of choice to reconstruct primordial tensor perturbations  is the eigen-analysis of the Fisher matrix of parameters that describe possible perturbations  to the primordial tensor spectrum (expanded in certain bases). 

As the fiducial model, we take the primordial tenor power spectrum, $P_{\rm t}^{\rm fid}(k)$ to have the standard power-law form of Eq.~\ref{fiducial},  with $k_{\rm t}=0.002 {\rm Mpc^{-1}}$ and $r=0.05$, unless explicitly stated otherwise (as in Section~\ref{sec:other}). 
The tensor tilt  is  also assumed to be vanishing, i.e., $n_{\rm t}=0$. The cosmological model is assumed to be the standard model of $\Lambda$CDM with the rest of standard parameters as reported for the baseline model of Planck 2018 \citep{pl18}. 

\begin{figure}
\begin{center}
\includegraphics[scale=0.35]{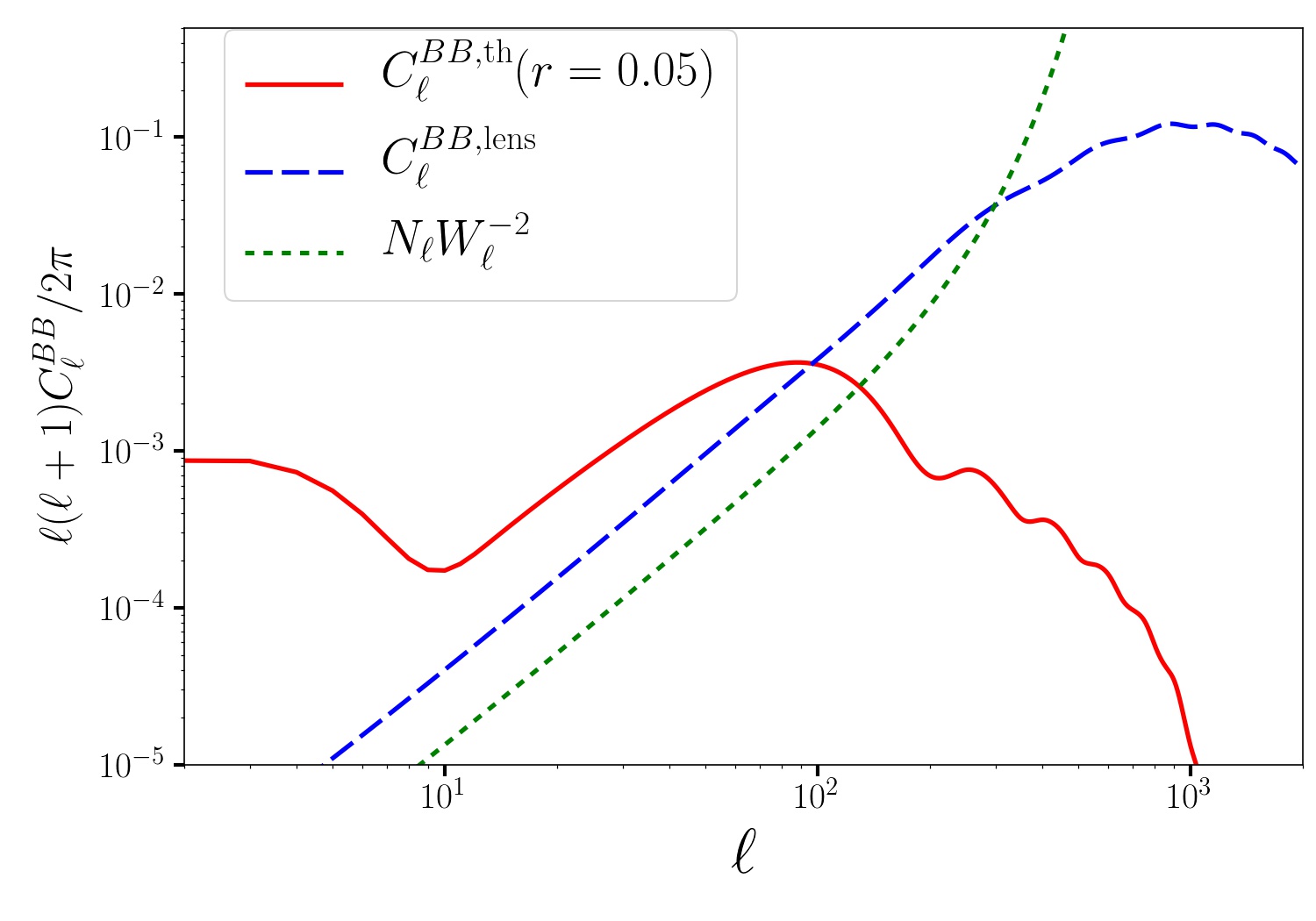}
\end{center}
\caption{Comparison of various contributions to the B-mode power spectrum. The experimental noise level is $\sigma_{\rm n}=0.1\mu K $ with a beam of $\theta_{\rm FWHM}=30'$.}
\label{fig:power}
\end{figure}

The polarization maps are also assumed to be partially delensed, with  the residual lensing contamination parametrized by $\lambda$. 
The observed B-mode power spectrum, $C_\ell^{BB}$,  would then be
\begin{equation}\label{eq:cl_obs}
C_\ell^{BB}=C_\ell^{BB, {\rm th}} W_\ell^2+N_\ell 
\end{equation}
where $W_\ell$ and $N_\ell$ describe the instrumental beam and noise respectively, and 
\begin{equation}\label{eq:cl_th}
 C_\ell^{BB,{\rm th}} =C_\ell^{ BB, {\rm prim}} + \lambda C_\ell^{ BB,{\rm  lens}}
\end{equation}
 with $C_\ell^{BB,{ \rm prim}} $ and $C_\ell^{BB, {\rm lens}}$ representing the primordial and lensing B-mode spectra.
Throughout this paper, we work with two observational setups. One case corresponds to a cosmic variance limited (CVL) experiment, where the effects of noise and beam are ignored. 
In the other case the anisotropies are smoothed by the instrumental beam, assumed be Guassian, $W_\ell \propto \exp(-\ell^2 \sigma_{\rm b}^2/2)$, where $\sigma_{\rm b}=0.425 \theta_{\rm FWHM}$ and ${\theta_{\rm FWHM}}=30'$ (e.g., corresponding to a Lite-bird-like experiment \citep{mat14}). We also assume a Gaussian white noise with $N_\ell= \sigma_{\rm n}^2\times {\rm pixel~area}$, where $\sigma_{\rm n}$ is the noise per pixel, here taken to be $0.1\mu K$.
In more general cases the noise term $N_\ell$ could have scale-dependent contributions as well, e.g., sourced by   foreground contamination and $1/f$-noise (see, e.g.,\citep{tho18,far13}) . We leave the treatment of $\ell$-dependent noise and their impact on the observability of the primordial signal to  a future work.
Fig.~\ref{fig:power} compares the various contributions (of Eq.~\ref{eq:cl_obs}) to the observed B-mode power spectrum, i.e., the primordial  spectrum sourced by gravitational waves (with $r=0.05$), the lensing spectrum and the experimental noise  (where for comparison, $N_\ell W_\ell^{-2}$ is plotted, with the specifications mentioned above.)
%-----------------------------------------------------------------------------------------------------------
%-----------------------------------------------------------------------------------------------------------
\begin{figure*}
\begin{center}
\includegraphics[scale=0.4]{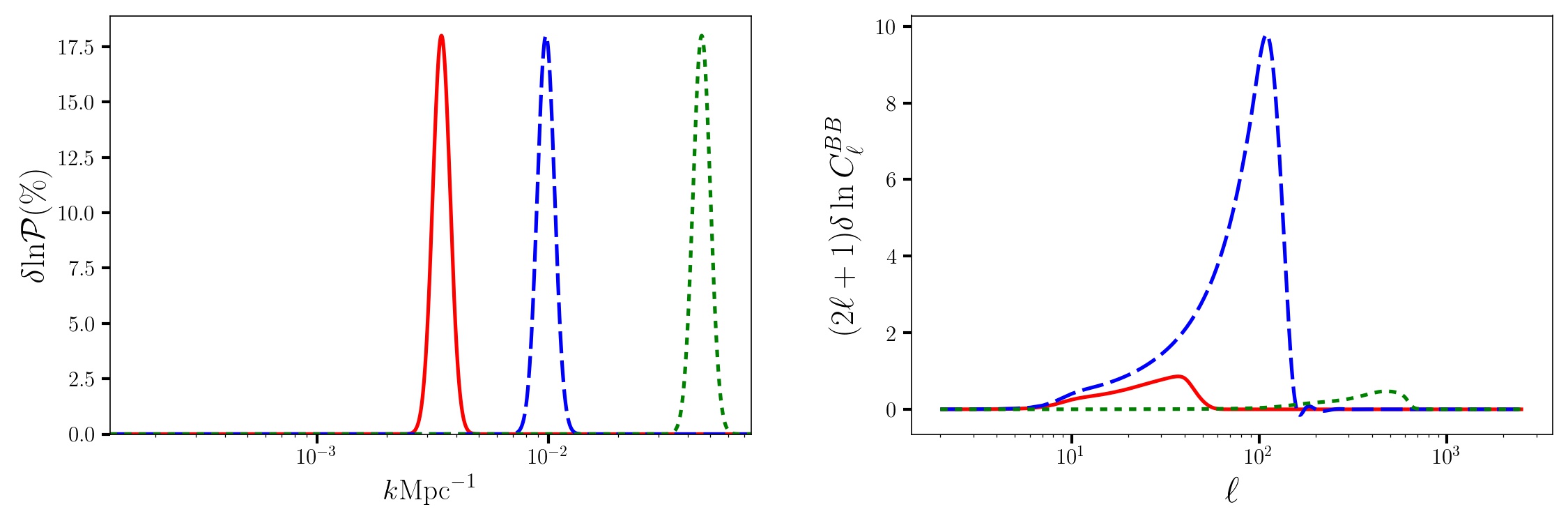}
\includegraphics[scale=0.4]{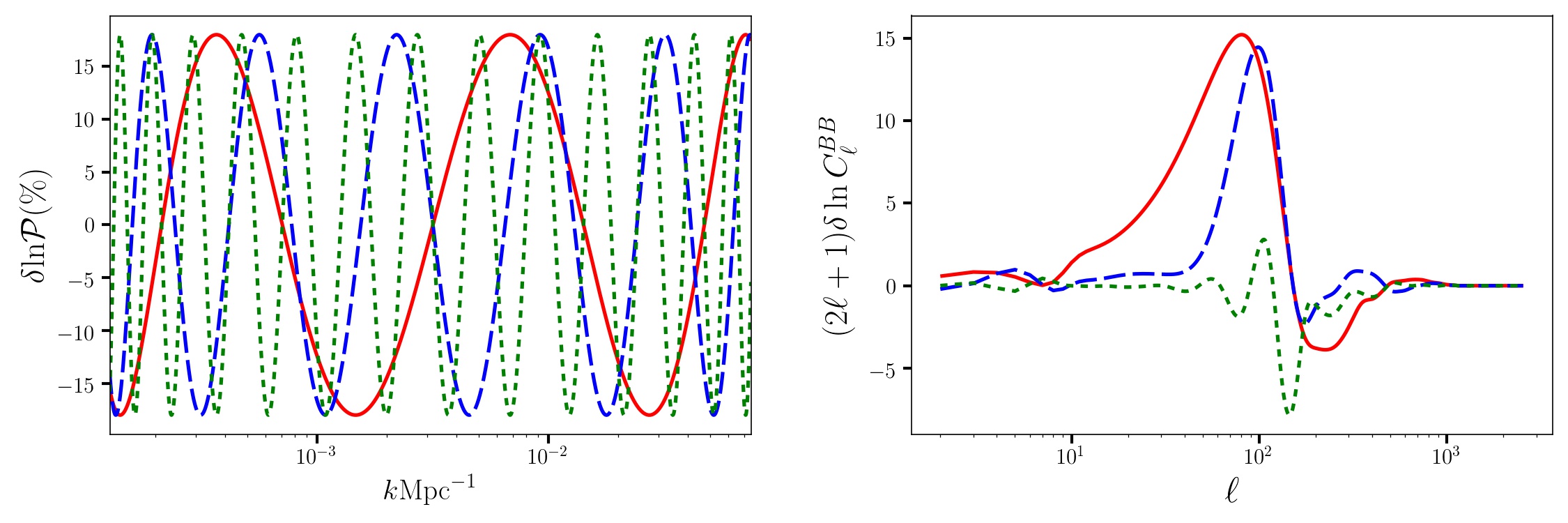}
\end{center}
\caption{Left: Examples of localized (top) and nonlocalized (bottom) basis functions in the form of Gaussian bumps and Chebyshev polynomials, respectively. Right: Their corresponding response in the B-mode power spectrum. The lensing contamination is assumed to be $50\%$ in these plots, i.e., $\lambda=0.5$ in Eq.~\ref{eq:cl_th}.}
\label{fig:basis}
\end{figure*}
%-----------------------------------------------------------------------------------------------------------

One can then relax the functional form of the tensor power spectrum and  allow for fluctuations around the fiducial pattern:
%-----------------------------------------------------------------------------------------------------------
\begin{equation}\label{eq:param}
\ln P_{\rm t}(k)=\ln P^{\rm fid}_{\rm t}(k)+\delta \ln P_{\rm t}(k)\\
%10^{-4} ~{\rm Mpc^{-1}} \le k \le 0.1 ~{\rm Mpc^{-1}}
\end{equation}
%-----------------------------------------------------------------------------------------------------------
with  $  10^{-4} ~{\rm Mpc^{-1}} \le k \le 0.1 ~{\rm Mpc^{-1}}$.
Given that  $P^{\rm fid}_{\rm t}(k)$ is scale-invariant ($n_{\rm t}=0$ in Eq.~\ref{fiducial}),
 the choice of $\delta \ln P_{\rm t}(k)$ versus $\delta P_{\rm t}(k)$ as the perturbation parameter can be simply regarded as a normalization preference. 
A general perturbation, as any other function,  can be expanded using any proper set of (ideally orthonormal) basis functions $\{ \phi_i\}$, and be parametrized by the expansion coefficients:
%--------------------------------------------------------------
\begin{equation}\label{eq:pert}
\delta \ln P_{\rm t}(k)=\sum_{i=1}^{N} q_i \phi_i(k) + r(k),
\end{equation}   
%--------------------------------------------------------------
where the residual $r(k)$ tends to zero as the number of basis functions, $N$, goes to infinity. In other words, the $\phi_i$'s form a complete set of basis functions  in the limit of $N\rightarrow \infty$. The parameter $q_i$ are the expansion parameters and represent the strength of the signal in the $i$-th basis function. They are obtained by direct projection of $\delta \ln P_{\rm t}(k)$ onto the $\phi_i$'s:
%--------------------------------------------------------------
\begin{equation}\label{eq:ortho}
q_i=\int_{k_{\rm min}}^{k_{\rm max}} \delta \ln P_{\rm t}(k)\phi_i(k) w(k) \rm{d}\ln k,
\end{equation}
%--------------------------------------------------------------
where $w(k)$ is the weight function which respect to which the $\psi_i(k)$'s are orthonormal:
%--------------------------------------------------------------
\begin{equation}
\int_{k_{\rm min}}^{k_{\rm max}} \phi_i(k) \phi_j(k) w(k) {\rm d}\ln k =\delta_{ij}.
\end{equation}
%------------------------------------------------------------------------------------------------------

In this work we have used both localized and nonlocalized sets as  basis functions, with the number of bases, $N$, ranging from $10$ to $320$. 
As the localized sets we used top hats, Gaussian and triangular bumps, considered as approximations to the Dirac $\delta$-function. The $i$-th localized basis function centered at $k_i$ and with the width $\sigma_i$ is given by:
%--------------------------------------------------------------
\begin{equation}
\phi_i(k)\propto \exp(-\frac{[k-k_i]^2}{2\sigma_i^2})
\end{equation}   
%--------------------------------------------------------------
for Guassian bumps, by 
%--------------------------------------------------------------
\begin{equation}
\phi_i(k)\propto  
\begin{cases}
    1-|k-k_i|/\sigma_i & \text{if $|k-k_i|<\sigma_i$},\\
    0 & \text{otherwise},
  \end{cases}
\end{equation}   
%--------------------------------------------------------------
for triangular bumps and by 
%--------------------------------------------------------------
\begin{equation}
\phi_i(k)\propto  
\begin{cases}
    1 & \text{if $|k-k_i|<\sigma_i$},\\
    0 & \text{otherwise},
  \end{cases}
\end{equation}   
%--------------------------------------------------------------
for top hats. 
We have taken  $\sigma_i=\Delta_k /N$ for the Gaussians and top-hats and $\sigma_i=1.5\Delta_k /N$ for triangular bumps, where $\Delta_k$ represents the width of the $k$-range of interest.
 
 As the nonlocalized basis functions, we used the Fourier series, 
 %--------------------------------------------------------------
\begin{eqnarray}
\phi_i(k)\propto  cos(i \pi y)~~~~~ i=0,1,2,...\\
\phi_i(k)\propto  sin(i \pi y)~~~~~ i=1,2,...\\
y=\frac{k-k_{\rm mid}}{\Delta_k}, ~~ k_{\rm mid}=(k_{\rm min}+k_{\rm max})/2.
\end{eqnarray}    
%--------------------------------------------------------------
 %
 as well as the Chebyshev polynomials of the first kind. Chebyshevs are constructed through the recursion formula $T_{i+1}(x)=2xT_i(x)-T_{i-1}(x)$, with the assumption that $T_0(x)=1$ and $T_1(x)=x$. The Chebyshev polynomials are orthogonal to one another with respect to the weight $w(x)=1/\sqrt{1-x^2}$ in the interval $[-1,1]$.  The weight function used for the orthogonality of the rest of basis functions used in this work is $w(x)=1$.

 For a detailed discussion on the properties of these basis functions and their relative advantages and shortcomings in representing perturbations see \cite{far12}.
Fig.~\ref{fig:basis} illustrates examples of localized and nonlocalized functions (top: Guassian bumps, bottom: Chebyshev polynomials respectively). 
The left panel presents the response of   $C_\ell^{BB,{\rm th}}$ (Eq.~\ref{eq:cl_th}) to these perturbations of the tensor spectrum.
We see that Guassian bumps with equal amplitudes placed at different wave-numbers lead to totally different B-mode response. A bump located at $k \sim 0.01 ~ {\rm Mpc}^{-1}$ leads to a very strong signal (at $\ell \sim 100$), whereas the response  to the bumps at the lower and higher $k$'s are hardly observable.
That is because of the cosmic variance domination at large scales and the lensing contamination at small scales (instrumental noise is not considered in these plots). 
For the Chebyshevs, on the other hand, there is contribution from all scales for any given Chebyshev function. However, for high frequency perturbations, the B-mode responses  due the adjacent ups and downs in 
$\delta \ln P_{\rm t}(k)$ partially cancel out, leading to suppressed response at high frequencies.

With the $N$ parameters ${\boldsymbol  q}=\{q_1,...,q_N\}$ characterizing the perturbations (Eq.~\ref{eq:pert}), one can construct $N$ uncorrelated parameters from the linear combinations of the $q_i$'s, based on the eigenmodes of their Fisher information matrix,
%------------------------------------------------------------------------------------------------------
\begin{equation}
F_{ij} \equiv - \bigg \langle \frac{\partial^2 \ln p_{\rm f}}{\partial q_i \partial q_j} \bigg \rangle.
\end{equation} 
%------------------------------------------------------------------------------------------------------
The brackets represent ensemble averaging over realizations of the CMB B-mode sky, and the partial derivatives must be calculated at the fiducial values of the parameters, here all being zero.
Also,  $p_{\rm f}\equiv p({\boldsymbol q}|d,\mathcal{T})$ is the Bayesian posterior probability of the parameters 
${\boldsymbol q}$, given the data $d$, in the theory space of  $\mathcal{T}$. In the language of Bayesian analysis, the posterior is an update on the prior probability of the parameters $p_i=p({\boldsymbol q}| \mathcal{T})$, when  data $d$ becomes available. This update is driven by the likelihood, $\mathcal{L}({\boldsymbol q}|d,\mathcal{T})\equiv p({d|\boldsymbol q},\mathcal{T})$ through the relation $p_{\rm f} =\mathcal{L}p_{\rm i}/\mathcal{E}$ where the evidence $\mathcal{E}\equiv p(d|\mathcal{T})$ can be considered  as a normalization factor.

Assuming a uniform prior on the parameters ${\boldsymbol q}$ would reduce the Fisher matrix to 
%------------------------------------------------------------------------------------------------------
\begin{equation}
F_{ij} \equiv - \bigg \langle \frac{\partial^2 \ln \mathcal{L}}{\partial q_i \partial q_j} \bigg \rangle. 
\end{equation} 
%------------------------------------------------------------------------------------------------------
If $f_{\rm sky}$ is the fraction of sky observed, and if the cut-sky-induced coupling between various modes is negligible,  the Fisher matrix further simplifies to 
%------------------------------------------------------------------------------------------------------
\begin{equation}\label{eq:fisher}
F_{ij}=f_{\rm sky}\sum^{\ell_{\rm max}}_{\ell=2}\frac{2\ell+1}{2}\frac{\partial C_\ell^{BB}}{\partial q_i}  \frac{\partial C_\ell^{BB}}{\partial q_j} /  (C_{\ell}^{BB})^2.
\end{equation} 
%------------------------------------------------------------------------------------------------------
%where the standard assumption of Gaussianity of the CMB signal (and noise) has also been applied.  
Throughout this work, we assume $f_{\rm sky}=1$. The upper limit of the summation, $\ell_{\rm max}$, is set by the smallest scale present in the data, generally determined by the experimental beam. 
 Clearly, for the specific parameters used in this work, the change in the parameters would only alter  $C_\ell^{BB,{\rm prim}}$. Therefore the signal contribution to the Fisher, i.e., the derivatives of $C_{\ell}^{BB}$ is only sourced by the primordial power spectrum, while the noise term (the denominator) has contribution from the primordial (fiducial) spectrum,  lensing and instrumental noise.
%-----------------------------------------------------------------------------------------------------------
\begin{figure*}
\begin{center}
\includegraphics[scale=0.33]{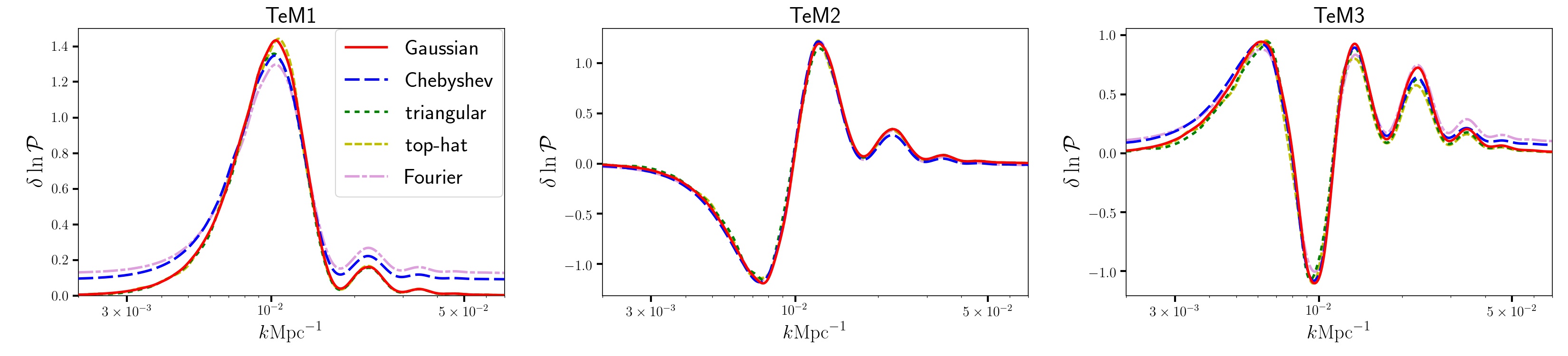}
\end{center}
\caption{The first three TeMs, constructed with the five different basis functions explained in the text with $N=160$,  for a CVL experiment and assuming the lensing residual is $50\%$ (case 2 of Table~\ref{table:expt}). }
\label{fig:basi-comp}
\end{figure*}
%-----------------------------------------------------------------------------------------------------------
\begin{figure*}
\begin{center}
\includegraphics[scale=0.33]{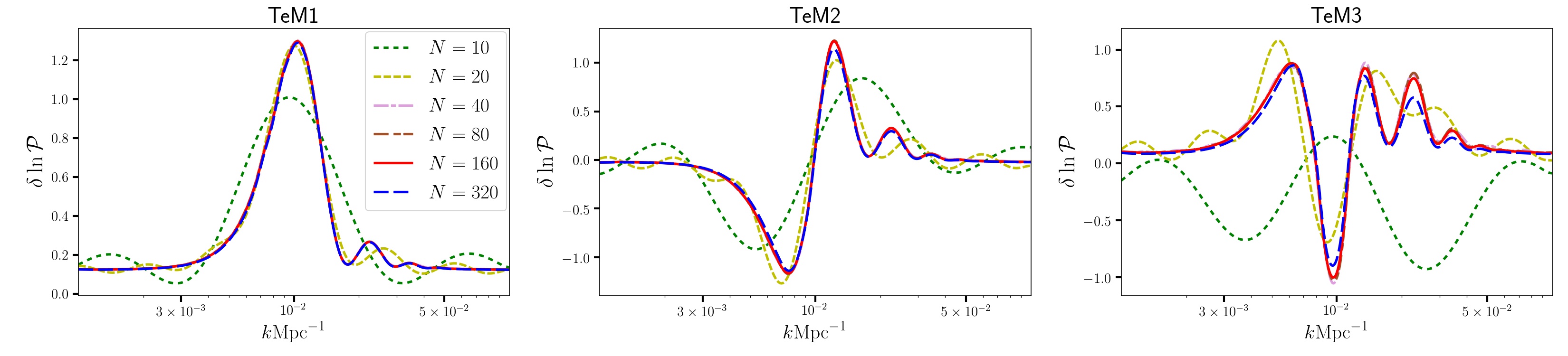}
\end{center}
\caption{Same as Fig.~\ref{fig:basi-comp}, constructed with the Gaussain basis function with various number of basis functions.}
\label{fig:Ncomp}
\end{figure*}
%-------------------------------------------------------------------------------------------

In the limit of a multi-variate Gaussian distribution for the parameters, the parameter correlation matrix 
$\boldsymbol{C}$ is approximated by the inverse of the Fisher matrix $C_{ij}=(F^{-1})_{ij}$.
In particular, the diagonal elements of the Fisher inverse correspond to the predicted errors for each parameter, 
$\sigma_i^2=(F^{-1})_{ii}$.

Unfortunately, these parameters are often weakly constrained and usually (highly) correlated. 
Fortunately, the Fisher matrix, as any $N\times N $ symmetric matrix, can be decomposed as $\boldsymbol{F}=\boldsymbol{S}^T \boldsymbol{f} \boldsymbol{S}$, 
where the columns of $\boldsymbol{S}$ are the $N$ (orthonormal) eigenvectors of $\boldsymbol{F}$ and $\boldsymbol{f}$ is a diagonal matrix whose diagonal elements are Fisher eigenvalues.

 One can therefor construct $N$ independent eigen-modes $E_j(k) ~~(j=1,..., N)$ from these eigenvectors and the basis functions we started with:
%--------------------------------------------------------------
 \begin{equation}
E_j(k)=\sum_{i=1}^{N} S_{ij}\phi_j(k) w(k)^{1/2}.
\end{equation}
%--------------------------------------------------------------
 The $ w(k)^{1/2}$ is included in the mode construction  to guarantee  orthogonality of the modes (see Eq.~\ref{eq:ortho}).
 These tensor eigenmodes, or TeMs hereafter, are by construction orthogonal  to each other (with unit weight, as can be easily verified) and normalized to one. 
They can therefore serve as a (complete) basis for the expansion of any perturbation in the primordial tensor power spectrum
 %--------------------------------------------------------------
\begin{equation}
\delta \ln P_{\rm t}(k)=\sum_{j=1}^{N} \mu_{j}E_j(k),
\end{equation}
%--------------------------------------------------------------
where 
%--------------------------------------------------------------
\begin{equation}
\mu_j=\int \delta \ln P_{\rm t}(k)E_j(k)\rm{d}\ln{k},
\end{equation}
%--------------------------------------------------------------
and we have assumed the residual of this expansion can be neglected. 
The  amplitudes of the TeMs ${\boldsymbol \mu}=\{\mu_1,...,\mu_N\}$ form a set of $N$ uncorrelated parameters capable of fully describing any perturbations that the original basis functions could cover.
They form  a hierarchy of  parameters in increasing order of estimated errors. One could thus limit the analysis to the most tightly constrained TeMs, as these represent the only practically observable part of the perturbation. 
The estimated uncertainty in the measurement of each $\mu_i$ is determined by the inverse square root of the corresponding eigenvalue, i.e., $f_{ij}=\sigma_i^{-2} \delta_{ij}$. In the case of non-Gaussian parameter distributions, the $\sigma$'s would give a lower bound on the parameter uncertainties. 
%-----------------------------------------------------------------------------------------------------------
\subsection{Tensor eigenmodes}\label{sec:tem}
%-----------------------------------------------------------------------------------------------------------
\subsubsection{Baseline scenario}
%-----------------------------------------------------------------------------------------------------------
Following the above procedure, we constructed the TeMs for the five  sets of basis functions introduced before (i.e., triangular and Gaussian bumps, top hats, Chebyshev polynomials and Fourier functions) for the baseline experimental scenario (case 2 of Table~\ref{table:expt}). It corresponds to a CVL full-sky observation, with $r_{\rm fid}=0.05$ and includes contamination from $50\%$ lensing residual.  
Fig.~\ref{fig:basi-comp} compares the first three TeMs constructed with these different basis sets and  clearly shows that the modes look practically the same irrespective of the basis choice. The plotted TeMs are also slightly smoothed to get rid of tiny wiggles sourced by numerical noise.
 We choose the Gaussian bumps as  the main basis for the rest of the work.

We also test the robustness of the TeMs against the number of basis functions, $N$, used for their construction. Fig.~\ref{fig:Ncomp} compares the first three TeMs for various $N$'s. We find that for $N=80$ the first three modes have already converged, and using  higher $N$'s does not practically affect the results. To make sure there is no convergence issue,  we choose $N=160$ as our choice of basis number in the rest of the work.\footnote{The TeMs for the various scenarios discussed here and the code for their generation, TEIMORE (standing for tensor eigenmode reconstruction), is straightforwardly applicable to other scenarios and is available upon request. TEIMORE uses the CMB B-mode spcetrum generated by the publically available Boltzmann code, CAMB (\url{http://camb.info}).}

% ,  is available at \mf{\url{http://Will be replaced after acceptance}} and is straightforwardly applicable to other scenarios}.  

Given the hierarchy of the modes, a practical  cut-off scheme is required to keep the most constrainable modes for the analysis of the data and discard the rest. There is no unique recipe for this and various criteria could be applied. 
For example, in the search for possible deviations from the standard recombination scenario,  \cite{far12}  used an information-based criterion for the  truncation of the eigenmode hierarchy.
 Here, by visual comparison, we note that for the scenarios considered in this work, TeM4  and higher get noisy and thus the first three TeMs are kept.

%-------------------------------------------------------------------------------------------
\begin{table}\centering
    \begin{tabular}{c|ccc}
    \toprule
obs. &  $r_{\rm fid}$ &  lensing residual &  $\sigma_{\rm n} $ \\ \hline 
%& & ($\%$ )& $(\mu {\rm  K arcmin})$ \\ \hline
         case 1 &  0.05 &  $10\%$ &   0 \\
     case 2 &  0.05 &  $50\%$ &   0 \\
     case 3 &   0.05 &  $10\%$ &   $ 0.1\mu {\rm  K}$ \\
      case 4 &  0.01 &   $50\%$ &   0 \\
    case 5 &  0.005 &   $10\%$ &   0 \\
     \bottomrule
\end{tabular}
\caption{The fiducial model and experimental specifications used for TeM generation as discussed in Section~\ref{sec:other}. $\sigma_{\rm n}$ is the noise per pixel, where the pixel side is determined by $\theta_{\rm FWHM}=30'$.}
\label{table:expt}
\end{table}
%-------------------------------------------------------------------------------------------
\begin{table}\centering%\begin{centering}
    \begin{tabular}{c|ccc}%{llll}
    %\hline
    \toprule
obs. &  $\sigma_1$ &  $\sigma_2$ &  $\sigma_3$ \\ \hline
     case 1 &  0.009 &  0.016 &  0.020 \\
     case 2 &   0.014 &   0.030 &  0.050 \\
     case 3 &   0.014 &  0.029 &  0.049 \\
     case 4 &   0.030 &  0.076 &   0.136 \\
     case 5 &   0.019 &   0.043 &  0.074 \\
     \bottomrule
\end{tabular}
\caption{Estimated uncertainities in the measurement of the first three TeMs (Fig.~\ref{fig:noiseffect}) for the five observational setups of Table~\ref{table:expt}, and discussed in Section~\ref{sec:other}.}
\label{table:errors}
\end{table}

%-----------------------------------------------------------------------------------------------------------
\subsubsection{Other scenarios}\label{sec:other}
\begin{figure*}
\begin{center}
\includegraphics[scale=0.33]{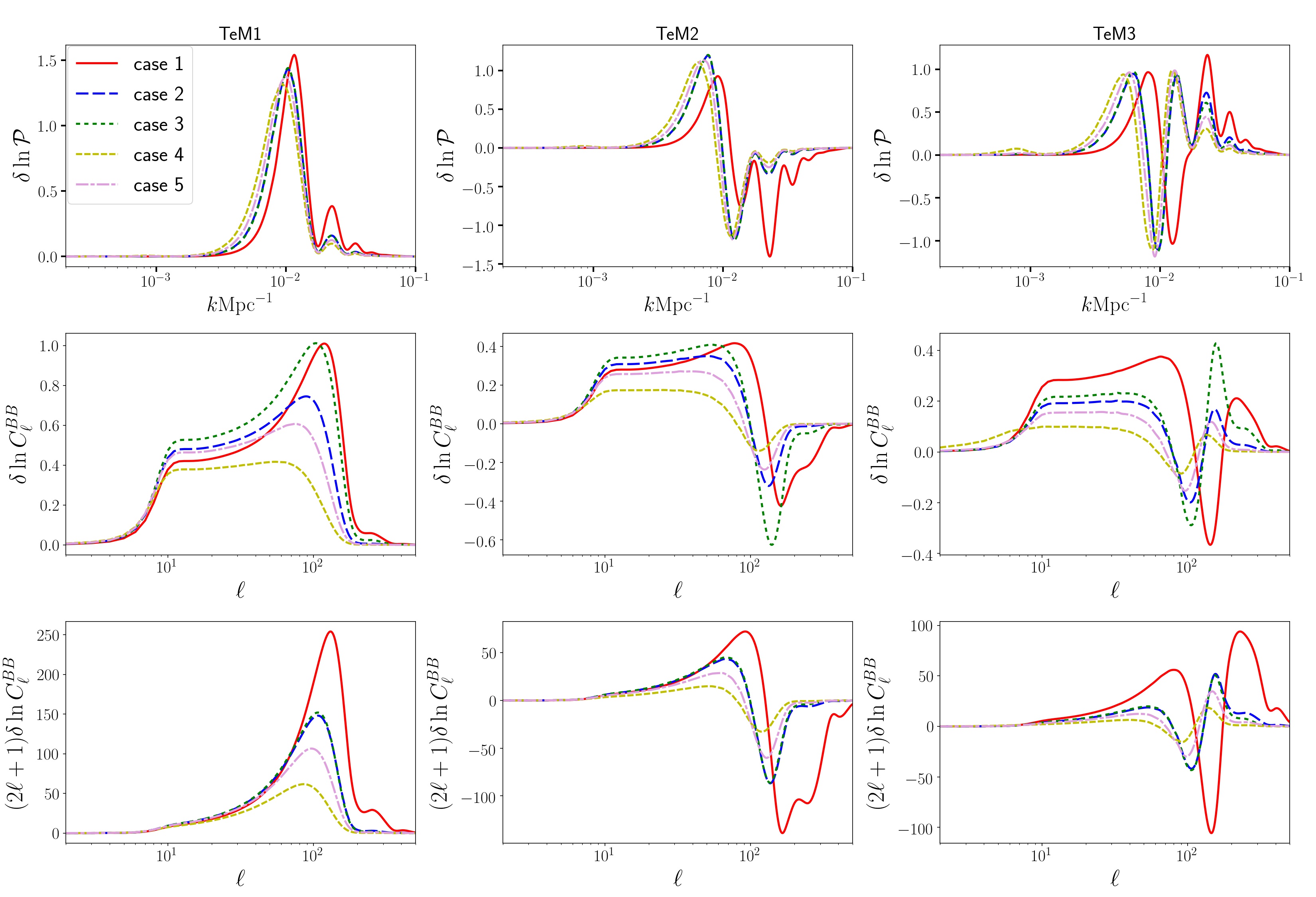}
\end{center}
\caption{Top: The first three TeMs, constructed with various assumptions for the fiducial $r$-value, noise and lensing contamination. See Table~\ref{table:expt}.  Middle: The response in the (theoretical) power spectrum 
$\delta \ln C_\ell^{BB,{\rm th}}$ to perturbations in the shape of the TeMs of the top row. Bottom: The SNR $(2\ell+1)\delta \ln C_\ell^{BB}$, as it goes in the construction of the Fisher matrix (Eq.~\ref{eq:fisher}) for different $\ell$'s corresponding to the TeMs of the top row. }
\label{fig:noiseffect}
\end{figure*}

%-----------------------------------------------------------------------------------------------------------
We investigated the sensitivity of the TeMs to the assumptions of the baseline scenario, and considered four more observational cases with different noise levels, fiducial $r$-values and lensing residuals, as listed in Table~\ref{table:expt}.
The first three TeMs for these cases, and their corresponding impact on the B-mode power spectrum are plotted in Fig.~\ref{fig:noiseffect}. The middle row is the relative change in the power spectrum due to the perturbation in the form of the associated TeM, and the bottom row illustrates the contribution to the Fisher (Eq.~\ref{eq:fisher}) as a function of $\ell$. 

We see that although the modes start to affect the power spectrum from $\ell\sim 10$ (middle row), it is the vicinity of $\ell \sim100$ that has the major contribution to the Fisher (bottom row) and therefor shapes the modes. That is due to the contamination from high cosmic variance at lower $\ell$'s and from  lensing (and noise, if presetn) at smaller scales.
Table~\ref{table:errors} presents the expected uncertainty in measuring the amplitudes of these three TeMs.
Note that the TeMs are $\delta\ln P(k)$ and not $\delta P(k)$, and therefore larger $r$'s would contribute both to the signal and the noise (i.e., to $\partial C_\ell^{BB}/ \partial q_j$ and $C_\ell^{BB}$ in Eq.~\ref{eq:fisher}, respectively).
Also note that increasing the noise (whether instrumental or lensing residual) or alternatively decreasing the signal (through lowering $r$) would shift the TeMs to lower $k$'s and damp them at higher $k$'s. 
That is because at lower SNRs, the noise would start to dominate the signal at larger scales (lower $k$'s) and thus the $C_\ell$'s lose their sensitivity to  perturbations at those scales.
%-------------------------------------------------------------------------------------------
%-----------------------------------------------------------------------------------------------------------

%----------------------------------------------------------------------
\subsection{Model reconstruction}\label{sec:models}
%----------------------------------------------------------------------------------------------------------
The TeMs, by construction,  form a set of (practically) orthonormal functions and can therefore serve as a complete basis for the expansion of any perturbation to the primordial tensor power spectrum.
However, it should be noted that very spiky features (compared to the highest frequency present in the nonlocalized functions, or the width of the localized basis functions) will not be  recoverable by the modes. Sharp rises at low or high wave-numbers are not also well reconstructed by the modes.
 Moreover, features requiring very high TeMs for their reconstruction should be considered practically 
 unconstrainable as the high eigenmodes are numerical-noise dominated.
These do not pose an issue, however.  They simply imply that the CMB B-mode polarization  is not sensitive to features at very high or very low $k$'s or to very spiky signals. 
In other words, in the absence of prior probabilities and theoretical prejudices, data do not recognize these features as observable.
Detection of such patterns requires a  devoted and biased search, as opposed to our (smei-)blind analysis, which can be justified by strong physical motivations. 

\begin{figure}
\begin{center}
\includegraphics[scale=0.5]{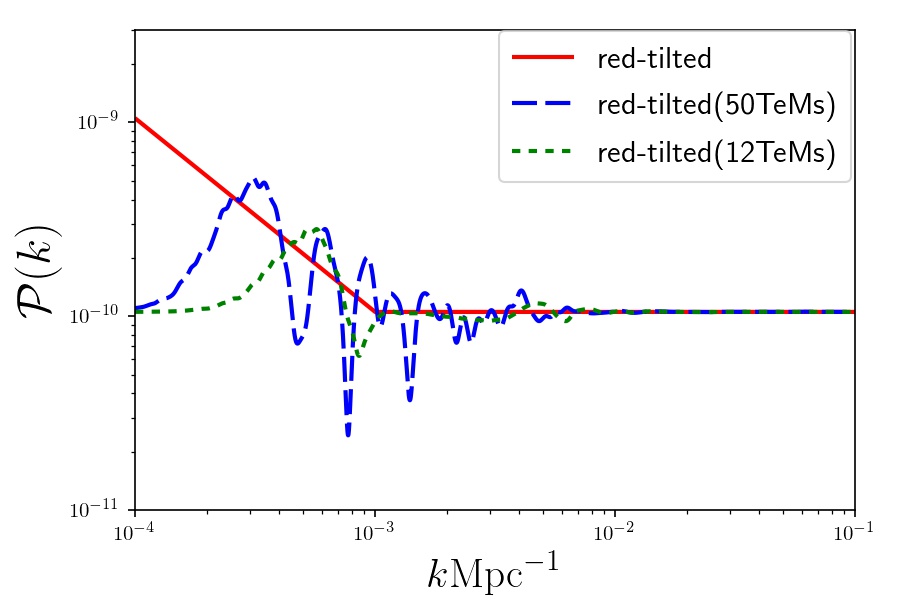}
\end{center}
\caption{A red-tilted primordial power spectrum (red solid line), and its reconstruction with the first twelve and the first fifty TeMs (green short-dashed and blue long-dashed lines, respectively).}
\label{fig:reconstruct}
\end{figure}

For demonstration purposes only, we use the TeMs to reconstruct a red-tilted power spectrum at large scales (present, e.g., in an open inflation scenario associated with a bubble nucleation, \cite{yam11}). We assume the power spectrum to coincide with our fiducial model at small scales, i.e.,   $P_{\rm t}(k)=P^{\rm fid}_{\rm t}(k)$ (with $n_{\rm t}=0$)  at $k \ge k_1$ and have a red spectrum at large scales, i.e., 
  $P_{\rm t}(k) \propto (k/k_1)^{n_{\rm t_1}}$ at $k < k_1$. We take $(k_1,n_{{\rm t}_1})=(0.001~{\rm Mpc}^{-1}, -1.0)$.
We find that fifty TeMs are needed for an acceptable recovery of the original power spectrum (Fig.~\ref{fig:reconstruct}) and yet the very large scales are not yet captured by the modes. The reconstruction of the the rise at the lowest $k$'s  requires a significant contribution of the higher modes, which are highly noisy. That is because the data is not sensitive to those scales and therefor the first few measurable TeMs, mainly localized at larger wavenumbers,  cannot mimic the large-scale rise. 

%-------------------------------------------------------------------------------------------
%-----------------------------------------------------------------------------------------------------------

It should also be stressed that the TeMs are by construction optimal for searches of small deviations around the assumed fiducial tensor power spectrum (here, a power-law with $r=0.05$ and $n_{\rm t}=0$ in Eq.~\ref{fiducial}).
 Power spectra significantly deviating from this model would be best reconstructed with TeMs generated around a different fiducial model, e.g., with a different $r$ and $n_{\rm t}$ corresponding to the best-fit values for that model (See Section~\ref{sec:disc} for more discussion).
The dependency of the modes on the fiducial model and experimental specifications was discussed in Section~\ref{sec:other} (see Fig.~\ref{fig:noiseffect}).

%--------------------------------------------------------------------------------------------
\section{Discussion}\label{sec:disc}
Different early universe scenarios predict different shapes for the primordial tensor perturbations.
Our goal in this work was to exhaust  the potential of the high precision B-mode polarization, in a model-independent way, to explore the physics of early universe through the reconstruction of the primordial tensor spectrum.
We therefor relaxed the functional form for the primordial tensor power spectrum to allow for small deviations around the widely-used and well-motivated power law.   We expanded possible deviations to this scenario using orthonormal basis functions and  applied eigen-analysis to the expansion coefficient in order to find the perturbation patterns to which the CMB B-mode is most sensitive We called these patterns tensor eigenmodes or TeMs. 
We checked for the robustness of the modes against the choice of basis functions used for  perturbation expansion as well as against  the number of basis functions used. We also investigated the sensitivity of the modes to the parameters of  fiducial model and the experimental assumptions. 

We claimed that the TeMs serve as a complete basis and any perturbation to the fiducial spectrum can be expanded in this basis. 
Patterns whose projections on the first few TeMs yield similar amplitudes are not distinguishable by data and are considered degenerate. That is because their difference mainly lies in the part of the perturbation space which is spanned by  higher and thus unconstrained TeMs. In other words, in a blind analysis, the only detectable parts of any perturbation are the ones with a significant projection onto the first few TeMs.

Moreover, if the power spectra of a given model is so far from the fiducial model that it can no longer be considered as a perturbation, we do not expect it  to be well reconstructed by the above TeMs.
 In such cases theoretical motivations could suggest different choices of the prior and/or the parametrization (Eq.~\ref{eq:param}) with more weight on scales where the main contribution of the signal is. Therefore, the resulting modes would be oriented to search for the possible fingerprints of the given model in the data.
These modes are clearly more appropriate for the reconstruction of perturbations close to the model of our interest. However, they are obtained at the cost of the partial loss of the blindness of the algorithm. 

In this work, our search for fluctuations in the primordial tenor spectrum was quite blind and only based on the simplest single-filed slow-roll inflationary models.
 Therefore, a uniform weight (in $\delta \ln k$), also supported by the almost scale-invariant power-law spectrum,  was assumed in our parametrization for different scales (Eq.~\ref{eq:param}). However, if there is good justification to search for, e.g., signals centered around a specific scale $k_*$, a more suitable parameter choice would be $ \delta \ln P(k)/(1+|k-k_*|/k_*)$. The choice of parametrization is obviously never unique and depends on the case. 
In the absence of theoretical priors, one could proceed as follows: first, assuming a fiducial power-law for the primordial spectrum, find the best-fit $(r,n_{\rm t})$ for the given data set. Then find the TeMs specific to this fiducial. 
As mentioned above, these modes have uniform prior weight in the $k$-range of interest and their features are solely due to the greater power of the data in constraining them. If we find out there is  significant power in the first few modes, we can include in our fiducial the reconstructed part of the perturbations using these TeMs  and repeat the process, i.e., generate the new TeMs specific to the new fiducial. 
 We terminate this iterative process of mode construction and fiducial-update when the amplitudes of the best-constrained modes are consistent with zero. One then starts the search for power in the higher (and noisy) modes. If there are hints for signals at these modes, we can regenerate the modes starting with a different parametrization weighted toward the scales with hints of signal. 
 The new TeMs should now be more suitable to locate those features and therefore less noisy at the desired scales. Further iteration on this procedure could continue until convergence on the TeMs is achieved and no more signal is recoverable from the data in this way.

\acknowledgements The numerical simulations were carried out on Baobab at the computing
cluster of the University of Geneva.

\bibliography{teimore}
\bibliographystyle{apj}

\end{document}